# Steady and Energy-Efficient Multi-Hop Clustering for Flying Ad-Hoc Networks (FANETs)

Basilis Mamalis
University of West Attica
Agiou Spyridonos, 12243,
Athens, Greece
vmamalis@uniwa.gr

Marios Perlitis
Democritus University of Thrace,
University Campus, 69100,
Komotini, Greece
mperlitis@gmail.com

## ABSTRACT
Flying Ad-hoc Networks (FANETs), formed by Unmanned Aerial Vehicles (UAVs), represent an emerging and promising communication paradigm. These networks face unique challenges due to UAVs' high mobility, limited energy resources, and dynamic topology. In this work, we propose a novel multi-hop clustering algorithm aimed at creating stable, energy-efficient clusters in FANET environments. The proposed solution enhances cluster longevity and communication efficiency through mobility-aware clustering, energy-centric cluster head (CH) selection, and a ground station(GS)-assisted cluster maintenance management mechanism. First, steady multi-hop clusters are constructed, having CHs with not only high stability and high energy but also with steady and high-energy neighboring areas, and then a proper GS-assisted cluster maintenance mechanism is applied. Experimental results, based on extended simulations, demonstrate that our approach outperforms existing schemes significantly, in terms of cluster stability, communication overhead, and security resilience.

## 1. INTRODUCTION
Unmanned aerial vehicles (UAVs), following the evolution of wireless technologies, are increasingly advancing towards clustering, networking, and broad intelligent operation [1,2]. Ensuring reliable and stable communication among UAVs is nowadays essential for coordinated swarm activities and continues to be a central area of research. In this context, the concept of the Flying Ad Hoc Network (FANET) [3,4] has emerged. In a FANET, the UAVs function as independent communication nodes, which allows them to autonomously establish a decentralized wireless network through mutual information exchange. Moreover, equipped with wireless communication modules and basic onboard sensors, UAVs can function collaboratively as a cohesive, networked group. FANETs are becoming increasingly prevalent and offer services that reduce the need for human involvement, especially in hazardous or hard to reach and practically inaccessible environments, thus minimizing potential risks to human life [5–9]. As a result, FANETs are attracting growing interest in the research community during the last decade [10].

Their adaptability, robustness, cost-efficiency, and ease of deployment make FANETs suitable for a wide range of applications—both civilian and military [11,12]. The concept of multi-UAV cooperation is central to FANETs and opens up possibilities in numerous fields, including high-precision geolocation [13], search and rescue operations [14], intelligent transportation systems [15], target recognition [12], disaster response and monitoring [17], volcano observation [18], and delivery of medical supplies to hard-to-reach areas [19]. They are also used in border surveillance [20], forest fire prevention and management [21], brain-controlled UAV operations [22], and as relay nodes for Internet distribution [23]. FANETs have become increasingly important in military operations as well. For instance, the U.S. Navy's LOCUST project employs swarms of autonomous drones to execute coordinated missions [25]. Beyond these established use cases, ongoing research and development in both academia and industry are exploring additional applications, such as surveying and mapping [26,27]. Through swarm coordination, UAVs can carry different sensors and operate in the same area concurrently, allowing for faster data collection and more efficient mission execution [28]. This collaborative behavior envisions a future where UAVs are seamlessly integrated into daily life, contributing to improved quality of life [29,30]. Despite these promising prospects, several technological challenges remain. These include rapidly changing network topologies, high node mobility, limited energy resources, reliable communication and coordination among UAVs, maintaining stable links between UAVs and ground stations, managing variations in transmission range and node density, and addressing critical security concerns in FANETs.

In this work, we propose a multi-hop clustering algorithm tailored to the aerial dynamics of UAVs. It aims to create steady clusters using mobility and energy similarity metrics while integrating ground station (GS)-assisted cluster maintenance mechanisms to enhance robustness and security. The proposed cluster formation algorithm initially constructs multi-hop clusters in which both the cluster heads (CHs) and the cluster members (CMs) located in proximity to the CHs exhibit high stability and substantial residual energy. Subsequently, an appropriate ground station (GS)-assisted maintenance mechanism is implemented to further enhance the stability and robustness of the established clusters. The remainder of the paper is structured as follows. Section 2 provides the necessary background, highlighting the recent advancements in FANET clustering. Section 3 presents a detailed description of the proposed clustering approach. In Section 4, our extended simulation experiments are introduced and a comprehensive discussion of the results is provided. Finally, Section 5 concludes the paper.

## 2. CLUSTERING IN FANETs

FANETs are characterized by high node mobility, rapidly changing topologies, and limited energy resources [3,31]. As the number of UAVs in a network grows, coordination and data exchange become more complex, increasing system instability. These challenges make designing stable and efficient FANET solutions difficult [32]. As illustrated in Fig. 1, a widely adopted approach is to divide the network into multiple subnets using clustering strategies. This hierarchical management mitigates the effects of local topology changes and enhances overall network performance in terms of stability, energy efficiency, and packet reception rate.

Clustering algorithms aim to form and maintain stable clusters at minimal computational and communication cost. The selection of cluster heads (CHs) is critical to the stability of the cluster structure. Over the years, numerous algorithms have been proposed, including the Highest Degree (HD) algorithm [33], the Lowest-ID (LID) algorithm [34], the Weighted Clustering Algorithm (WCA) [35], and the Mobility-Based Clustering (MOBIC) [36]. While the other three rely on single factors and struggle to adapt to dynamic environments, WCA incorporates multiple parameters—such as node position, mobility, and energy consumption—allowing it to adapt better to various application scenarios. As a result, many modern clustering algorithms (even among the latest) are enhancements of WCA.

Depending on their design goals, existing clustering approaches can be categorized into four types: mobility-based, trust-based, energy-based, and task-oriented. Mobility-based algorithms focus on adapting to dynamic topologies by tracking node movement [37]. Trust-based algorithms improve network security by identifying and isolating malicious or uncooperative nodes [38]. Energy-based methods aim to reduce energy consumption and extend network lifespan [39]. Task-oriented clustering is tailored to specific functional requirements or missions [40].

The effectiveness of clustering in FANETs largely depends on the mechanisms for selecting and switching cluster heads (CHs). Several strategies have been introduced. For instance, Zang et al. [37] proposed a link expiration time-based algorithm that uses location and mobility data. Shu et al [41] designed a mobility prediction-based routing algorithm for high-speed nodes, selecting CHs based on link stability with one-hop neighbors. Wang et al. [42] introduced a neural network approach based on gray wavelets and adaptive node degree. Singh et al. [38] introduced a trust-based CH selection method using fuzzy inference and a reward-punishment mechanism to enhance secure communication.

To improve cluster stability, researchers have also explored CH switching and backup mechanisms. Pathak et al. [43] reduced CH changes and overhead by incorporating backup nodes prioritized by node degree and remaining energy. Mei et al. [44] and Wang et al. [45] proposed dynamic weight assignment methods to increase algorithm adaptability. Aissa et al. [46] introduced safety distance considerations to select both CHs and backups, enhancing stability and energy efficiency. Also, Bhandari et al. [32] proposed a periodic maintenance mechanism based on node mobility, while Guo et al. [47] developed a low-latency maintenance scheme to mitigate delays caused by unexpected CH failures.

Beyond traditional methods, bio-inspired algorithms—drawn from natural swarm behaviors—have shown promise due to their distributed, adaptive, and resilient characteristics. Yu et al. [48] developed a clustering algorithm inspired by slime mold foraging behavior, leveraging UAV mobility patterns. Li et al. [49] combined an improved IK-means algorithm with an artificial bee colony approach for fast CH selection in dynamic environments. Aftab et al. [50] proposed a hybrid scheme using either firefly or dragonfly algorithms for effective cluster management. Liu et al. [51] proposed an adaptive algorithm (hummingbird-based meta-heuristic) that optimizes network topology with varying channels.

Other efforts have utilized metaheuristics like moth flame optimization [52,53] for cluster maintenance and construction. To overcome stability problems caused by high mobility and limited energy, Khan et al. [54] proposed a firefly-shrimp swarm hybrid clustering method. Zhang et al. [40] applied a gray wolf-based approach, integrating energy detection for CH selection. Arafat and Moh [55] used a 3D PSO-based algorithm to improve CH selection efficiency. Finally, Yan et al. [39] calculated the optimal number of clusters using bandwidth and coverage constraints, and then applied a binary whale optimization method for CH selection, with the aim, among other goals, of reducing energy consumption.

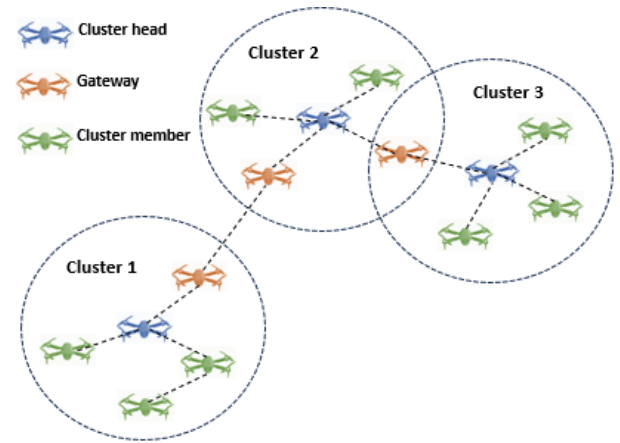

**Fig 1.** Cluster-based organization and routing in FANETs

## 3. THE PROPOSED MULTI-HOP CLUSTERING APPROACH

As described earlier, the proposed solution is built upon a clustering framework in which certain nodes—elected as cluster heads (CHs)—serve as leader UAVs for data communication and coordination tasks. These CHs either collect and forward data received from other UAVs within the cluster (members, CMs) or disseminate information to them. Data exchange occurs in a multi-hop fashion for both inter-cluster and intra-cluster communication. Thus, to efficiently manage overall cluster stability while maintaining balanced communication overhead, we employ a novel multi-hop clustering strategy. The proposed algorithm prioritizes cluster formation based on two key criteria: the overall stability and the spatial centrality of each UAV within its neighborhood. This design results in stable clusters with extended lifetimes and improved reliability. The proposed approach (named SEFC – Steady Energy Efficient Clustering) is developed under the following foundational assumptions.

### 3.1 Network Assumptions

- Each UAV has a unique ID, GPS, inertial navigation system (INS), and onboard processing.

- UAVs are equipped with 3D mobility tracking modules and limited energy resources.

- A ground station (GS), analogous to the RSU in VANETs, may be intermittently available and can assist in clusters maintenance.
- Aerial communication relies on IEEE 802.11s or similar mesh protocols.

### 3.2 Cluster Formation Steps

The basic steps of the proposed protocol for the formation of the clusters directly follow:

**i. UAV Advertisement:**

Each UAV broadcasts a beacon message containing its ID, position in the space (x, y, z), velocity vector, acceleration, and residual energy.

**ii. Mobility & Energy Similarity Evaluation:**

Each UAV $i$ calculates a *Mobility-Energy Difference (MED)* with its one-hop neighbors. Specifically, the above (*MED*) value is computed with respect to each neighbor node/UAV $j$ based on the differences between the two UAVs in *speed*, *acceleration* and *energy*, which are the most important factors giving a notion of the actual similarity in the mobility and resource behavior of both UAVs. More concretely, the corresponding differences between the two UAVs ($SD(i,j)$ $AD(i,j)$ and $ED(i,j)$ respectively) are computed (and then also normalized appropriately) as follows; where $S(i)$, $S(j)$, $A(i)$, $A(j)$, $E(i)$, $E(j)$ denote the speed, the acceleration and the residual energy of each UAV, and $SD_{max}(i)$, $AD_{max}(i)$, $ED_{max}(i)$ are the *maximum* speed, acceleration and residual energy differences among $i$ and all its neighbors:

$$SD(i,j) = \frac{|S(i) - S(j)|}{SD_{max}(i)}$$

$$AD(i,j) = \frac{|A(i) - A(j)|}{AD_{max}(i)}$$

$$ED(i,j) = \frac{E(j) - E(i)}{ED_{max}(i)}$$

Note also here that with respect to $SD(i,j)$ and $AD(i,j)$ it's sufficient to take in account the absolute values of the corresponding differences, *w*hereas with respect to $ED(i,j)$ we have to take in account the actual difference (in residual energy) of the nodes $j$ having only greater residual energy than $i$. The final *MED* value for each neighbor UAV $j$ with respect to $i$ is computed by the following formula, where $c1$ and $c2$ are (appropriately selected) constant coefficients, with $c1 + c2 + c3 = 1$.

$$MED(i,j) = c_1 \cdot SD(i,j) + c_2 \cdot AD(i,j) + c_3 \cdot (1 - ED(i,j))$$

Only neighbors below a *MED* threshold and flying in similar direction are retained as candidates for cluster membership. Note also here that, actually, the complement of $ED(i,j)$ ($1-ED(i,j)$) is taken in account in the above formula since we are interested to keep in the list the neighbors with sufficiently larger residual energy.

**iii. Overall Stability Factor (OSF) Calculation:**

Next, each UAV computes its *Overall Stability Factor (OSF)* based (a) on its total/average mobility (speed, acceleration) and residual energy differences with all the eligible (according to the computed above *MED* value) neighbors, as well as (b) on the degree (# of neighbors / node density) of the UAV node and its total/average distance form all its eligible neighbors. More concretely, the *OSF* value of each UAV $i$ is computed as follows.

$$OSF(i) = \alpha \cdot SD_{av}(i) + \beta \cdot AD_{av}(i) + \gamma \cdot ED_{av}(i) + \delta \cdot D_{av}(i) + \varepsilon \cdot d(i)$$

With respect to the above calculation, $\alpha$, $\beta$, $\gamma$, $\delta$ and $\varepsilon$ are (suitably selected) constant coefficients, with $\alpha + \beta + \gamma + \delta + \varepsilon = 1$, $d(i)$ stands as the degree of UAV node $i$, and $SD_{av}(i)$, $AD_{av}(i)$, $ED_{av}(i)$, $D_{av}(i)$ represent the justified measurements with regard to the average speed difference, the average acceleration difference, the average residual energy difference and the average relative distance, respectively, between UAV $i$ and all its neighbor UAVs (let's name it set $N_i$). The $SD_{av}(i)$, $AD_{av}(i)$, $ED_{av}(i)$ and $D_{av}(i)$ measures are more specifically calculated as follows.

$$SD_{av}(i) = \frac{\sum_{j \in N_i}(1 - SD(i,j))}{d(i)}$$

$$AD_{av}(i) = \frac{\sum_{j \in N_i}(1 - AD(i,j))}{d(i)}$$

$$ED_{av}(i) = \frac{\sum_{j \in N_i} ED(i,j)}{d(i)}$$

$$D_{av}(i) = \frac{1}{d(i)} \cdot \sum_{j \in N_i}(1 - \frac{D(i,j)}{D_{max}(i)})$$

where:

$$D(i,j) = \sqrt{(x_i - x_j)^2 + (y_i - y_j)^2 + (z_i - z_j)^2}$$

Note that $D_{max}(i)$ is the maximum distance of UAV $i$ from all its neighbors. Note also that all except $ED_{av}(i)$ are taken in account with suitable reverse normalization, in order to choose the maximum as the best.

Moreover, $D_{av}(i)$ – which represents the average distance value for each UAV – is computed as the cumulative mean square distance of the UAV to its direct neighbors, divided by its degree as shown above (considering that $x_i, y_i, z_i$ and $x_j, y_j, z_j$ are the location coordinates of UAVs $i$ and $j$ respectively).

**iv. UAV Stability (OSF) Broadcast:**

Each UAV broadcasts its OSF and ID.

**v. Parent Selection & Cluster Head (CH) Election:**

Each UAV selects the neighbor from its similarity set with the highest OSF (greater than its own) as its *parent*. If no suitable parent exists, it elects itself as a *Cluster Head (CH)*.

**vi. Backup Cluster Head (BKCH):**

The one-hop neighbors of the current *CHs* with the next-highest OSF value are designated as *backup CHs* (*BKCHs* – utilized later in cluster maintenance phase).

### 3.3 Cluster Maintenance Management

When the CH/CMs fall within the range of a Ground Station (GS), the following steps are executed:

**i. Local Stability Re-evaluation:**

The GS recalculates the OSF values based only on current cluster members, focusing on relative aerial movement and remaining energy.

**ii. CH Handover:**

If another node (e.g., the *BKCH*) has a significantly higher OSF, the *CH* role is reassigned accordingly.

**iii. Re-clustering Trigger:**

If the average OSF values fall below a threshold, the GS issues a re-clustering command and notifies accordingly the in-range *CHs* and *CMs*.

## 4. SIMULATION EXPERIMENTS AND PERFORMANCE ANALYSIS

In this section, we present in details the extended simulation experiments conducted to demonstrate the efficiency and scalability of the proposed approach. Specifically, we perform a comparative performance analysis between the proposed SEFC algorithm and two existing clustering algorithms— PICA [1] and OSCA [43]—using the OMNeT++ network simulator under identical simulation conditions.

The PICA algorithm was constructed to overcome issues related to the delay and stability of the network, considering UAV scenarios with high mobility. By considering UAV movement and residual energy, and using improved mobility and distance security metrics, PICA keeps drones at safe distances, maintains stable links and ensures longer network lifetime. Furthermore, it strengthens cluster maintenance through a backup CH selection mechanism and incorporates damage detection and cluster merging strategies to improve cluster robustness and stability.

Conversely, the OSCA algorithm focuses on enhancing network stability by minimizing cluster head changes and reducing clustering overhead. It introduces a backup node that takes over as the new cluster head when the current one fails or leaves the cluster, subsequently selecting a new backup node. This mechanism maintains network availability and limits interference. Additionally, OSCA determines the priorities of CHs and backup cluster-head nodes, based on node degree and residual battery life, thereby selecting the most suitable nodes for these roles.

The reader may easily notice that all the above three algorithms – PICA, OSCA, and the proposed SEFC algorithm – share the common goal of improving clustering stability and maintaining network availability through the backup cluster head concept. Apparently, this shared characteristic makes them suitable candidates for both technical comparison and reliable experimental evaluation.

Moreover, in the overall experimental design presented throughout this section, we've appropriately varied the number of nodes and the maximum node mobility speed to assess each algorithm's performance and robustness across different network scales and mobility conditions. Specifically, adjusting the number of nodes evaluates scalability and the load on cluster head nodes, while varying mobility speed and acceleration simulates realistic network dynamics, allowing analysis of communication delay and connection stability under different movement rates.

Finally, to reliably evaluate the clustering performance of the involved algorithms, we used four key metrics: average end-to-end delay, energy consumption, average cluster head duration, and the number of inter-cluster member (CM) switches. These metrics collectively assess communication performance, energy efficiency, and clustering stability, providing a comprehensive performance evaluation. This validation framework enables optimization and refinement of the clustering algorithm, enhancing its adaptability and reliability in real-world applications. The specific simulation parameters and their corresponding values are consistent with those used in [1], ensuring the reliability and comparability of the experimental results and conclusions.

### 4.1 Average end-to-end delay

The average end-to-end delay represents the mean time required for data packets to travel from source nodes to destination nodes. This metric is directly influenced by the number of UAVs in the network, as higher node density and mobility can lead to network congestion, packet loss, and consequently longer transmission times. When a clustering algorithm establishes a more stable network topology, communication between nodes becomes faster and more reliable, thereby reducing the overall end-to-end delay.

Figure 2 presents a comparative experimental analysis of the average end-to-end delay achieved by the proposed SEFC algorithm and the PICA and OSCA algorithms. The simulated experiments were performed by varying the number of UAVs from 40 to 140, with a maximum flight speed of 60m/s.

Among all simulation scenarios, we can notice that the SEFC algorithm steadily achieves the lowest average end-to-end delay, showing reductions of 11.6% and 24.3% compared to PICA and OSCA, respectively. As we can easily realize, this improvement is mainly based on the fact that SEFC enhances link stability via the weighted evaluation of multiple complementary criteria during CH selection. Moreover, by minimizing packet retransmissions, SEFC effectively reduces the overall delay. In contrast, PICA and OSCA primarily rely on a limited set of parameters, such as inter-node distance and relative mobility; a weaker approach that may lead to unstable transmission paths, higher number of packet retransmissions and (as a result) increased average end-to-end delay.

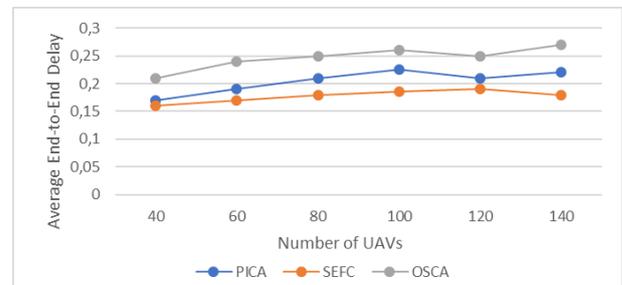

**Fig 2.** Average end-to-end delay time for varying # of UAVs

### 4.2 Average energy consumption

The energy consumption of UAVs serves as a crucial metric for evaluating the energy efficiency of clustering algorithms. As a consequence, an effective clustering approach should be capable of forming clusters that enable UAVs to communicate while minimizing energy expenditure. More concretely, lower levels of energy consumption show that the algorithm utilizes node resources efficiently during execution, thus prolonging the lifetime of the node and enhancing the stability of the system. Furthermore, reduced levels in energy consumption, also usually imply less communication overhead, which leads to reduced delays in data transmission, less bandwidth usage and increased network robustness. As a result, the lifetime of the network is prolonged, leading to more viable underlying applications. Figure 3 shows the results obtained with regard to energy consumption for varying numbers of UAVs.

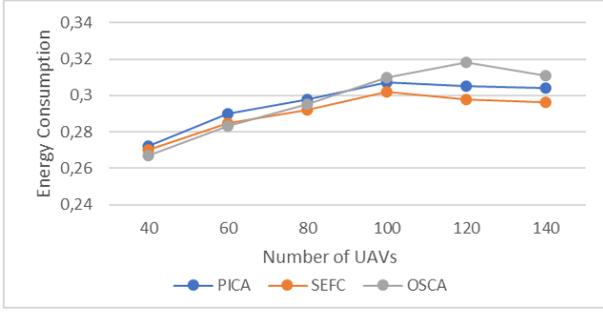

**Fig 3.** Average energy consumption for varying # of UAVs

As we can see in the above figure, SEFC (which employs a multi-criteria and multi-hop clustering approach) exhibits a slightly higher control overhead than PICA and OSCA when the number of UAVs is relatively small. This leads to a marginal increase in energy consumption of approximately 0.55% compared to OSCA. However, as the number of nodes grows, the multi-criteria basis of SEFC enhances network stability, resulting in energy consumption reductions of 2.45% and 3.65% relative to PICA and OSCA, respectively.

### 4.3 Average CH duration

Cluster head duration refers to the period during which a UAV maintains its role as a cluster head before transitioning to a non-cluster head state. Figure 4 presents a comparison of the SEFC, PICA, and OSCA algorithms in terms of average cluster head duration for a network of 100 UAVs operating at different maximum flight speeds. As UAV speed increases, the average cluster head retention time decreases because higher mobility accelerates topology changes, leading to more frequent disconnections between nodes. A longer average CH duration indicates that the clustering algorithm forms more stable cluster partitions capable of maintaining structural consistency even under dynamic network conditions. As we can easily realize, this leads to increased adaptability and potential to retain stable cluster structures. UAVs that keep the CH role for extended periods can provide more reliable communication and routing services, thus improving the overall network stability and performance.

Based on Figure 4, we can notice that the SEFC algorithm achieves (compared to PICA and OSCA) significantly longer cluster head retention times, with improvements of 22.18% and 83.27%, respectively. This superior performance results from SEFC's multi-criteria clustering mechanism, which combines multiple mobility-related factors during cluster head selection, thus enhancing cluster stability. In contrast, PICA and OSCA rely on less effective selection strategies, which cause frequent CH changes in highly dynamic environments due to constant node switching between clusters.

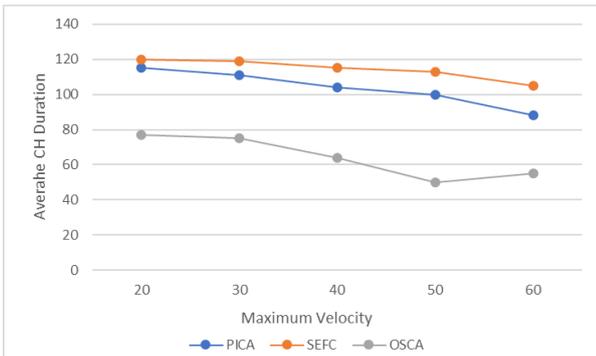

**Fig 4.** Average CH duration for varying maximum velocity

### 4.4 Average number of CM switches

The node cluster switching count measures cluster stability from the perspective of cluster members (CMs). A lower switching count indicates greater cluster stability. As shown in Figure 5, increasing node speed increases the potential for network topology changes, thus leading to higher switching counts across all the three involved algorithms.

A smaller average number of inter-cluster switches indicates that the algorithm effectively maintains sufficiently stable cluster structures despite the node mobility and network fluctuations, thus minimizing frequent cluster reassignments. This demonstrates the algorithm's adaptability and its ability to preserve consistent clustering results even under intense dynamic conditions. Additionally, fewer inter-cluster switches help reduce overall network overhead and communication delay, improving system efficiency.

Observing Figure 5, we can notice that SEFC achieves (compared to PICA and OSCA) reductions in the number of CM switches of 7.37% and 14.29%, respectively. This is mainly based on SEFC's multi-criteria clustering mechanism combined with its multi-hop organization structure, which together minimize frequent CM disconnections. In contrast, PICA and OSCA employ less effective clustering strategies, resulting in more frequent disconnections in highly dynamic environments. As a result, disconnected nodes must rejoin new clusters, incurring additional switching overhead.

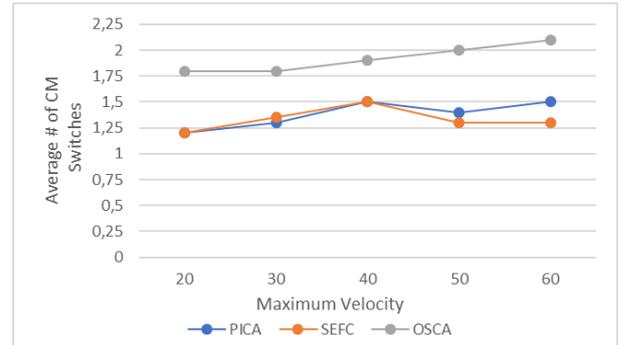

**Fig 5.** Average # of CM switches for varying maximum velocity

### 5. CONCLUSION

Throughout the present work, a novel multi-hop clustering algorithm tailored to the aerial dynamics of UAVs is presented and evaluated via simulated experiments. It aims to create steady clusters using multi-objective criteria, such as mobility and energy similarity metrics etc., while integrating ground station (GS)-assisted cluster maintenance mechanisms to enhance robustness and stability. The proposed cluster formation algorithm initially establishes multi-hop clusters in which both the cluster heads (CHs) and the cluster members (CMs) located near the CHs possess high stability and substantial residual energy. Subsequently, an appropriate GS-assisted maintenance mechanism is employed to further enhance the stability and longevity of the formed clusters. Simulation results demonstrate that the proposed approach outperforms existing schemes in terms of cluster stability, energy consumption, communication overhead, and end-to-end delay. The extension and optimization of the proposed algorithm for even more dynamic environments with real-time processing and re-action requirements, as well as for efficient involvement in modern edge computing scenarios for task offloading procedures, is of high priority in our future work.